\RequirePackage[left,columnwise]{lineno}
\documentclass[aps,prc,twocolumn,floatfix,superscriptaddress,nofootinbib]{revtex4}
\usepackage{lipsum}
\usepackage{graphicx}
\usepackage{epstopdf}
\usepackage{subfigure}
\usepackage{epsfig}
\usepackage{amsmath,amssymb,amsfonts,eufrak,mathrsfs}
\usepackage{bm}
\usepackage{color}
\usepackage[bookmarks,
                bookmarksopen = true,
                bookmarksnumbered = true,
                linktocpage,
                colorlinks = true,
                linkcolor = blue,
                urlcolor  = blue,
                citecolor = blue,
                anchorcolor = green,
                hyperindex = true,
                hyperfigures]
                {hyperref}

\usepackage{lineno}

\begin{document}

\setrunninglinenumbers
\setlength\linenumbersep{0.10cm}

\title{Multistage Monte-Carlo simulation of jet modification in a static medium}

\author{S. Cao\footnote{Corresponding Authors}}
\affiliation{Department of Physics and Astronomy, Wayne State University, Detroit MI 48201.}
\author{C. Park$^*$}
\affiliation{Department of Physics and Astronomy, McGill University, Montr\'{e}al QC H3A-2T8.}

\author{R.~A.~Barbieri}
\affiliation{Department of Physics, MIT, Cambridge MA 02139.}

\author{S.~A.~Bass}
\affiliation{Department of Physics, Duke University, Durham, NC 27708.}

\author{D.~Bazow}
\affiliation{Department of Physics, The Ohio State University, Columbus OH 43210.}

\author{J.~Bernhard}
\affiliation{Department of Physics, Duke University, Durham, NC 27708.}

\author{J.~Coleman}
\affiliation{Department of Statistics, Duke University, Durham, NC 27708.}

\author{R.~Fries}
\affiliation{Department of Physics and Astronomy, Texas A\&M University, College Station TX 77843.}
\affiliation{Cyclotron Institute, Texas A\&M University, College Station TX 77843.}

\author{C.~Gale}
\affiliation{Department of Physics and Astronomy, McGill University, Montr\'{e}al QC H3A-2T8.}

\author{Y.~He}
\affiliation{Key Laboratory of Quark and Lepton Physics (MOE) and Institute of Particle Physics, Central China Normal University, Wuhan 430079, China.}
\affiliation{Nuclear Science Division, Lawrence Berkeley National Laboratory, Berkeley CA 94720.}

\author{U.~Heinz}
\affiliation{Department of Physics, The Ohio State University, Columbus OH 43210.}

\author{B.~V.~Jacak}
\affiliation{Department of Physics, University of California, Berkeley CA 94720.}
\affiliation{Nuclear Science Division, Lawrence Berkeley National Laboratory, Berkeley CA 94720.}

\author{P.~M.~Jacobs}
\affiliation{Nuclear Science Division, Lawrence Berkeley National Laboratory, Berkeley CA 94720.}

\author{S.~Jeon}
\affiliation{Department of Physics and Astronomy, McGill University, Montr\'{e}al QC H3A-2T8.}

\author{M.~Kordell II}
\affiliation{Department of Physics and Astronomy, Wayne State University, Detroit MI 48201.}
\affiliation{Cyclotron Institute, Texas A\&M University, College Station TX 77843.}

\author{A.~Kumar}
\affiliation{Department of Physics and Astronomy, Wayne State University, Detroit MI 48201.}

\author{T.~Luo}
\affiliation{Key Laboratory of Quark and Lepton Physics (MOE) and Institute of Particle Physics, Central China Normal University, Wuhan 430079, China.}

\author{A.~Majumder}
\affiliation{Department of Physics and Astronomy, Wayne State University, Detroit MI 48201.}

\author{Y.~Nejahi}
\affiliation{Department of Computer Science, Wayne State University, Detroit MI 48202.}

\author{D.~Pablos}
\affiliation{Department of Physics and Astronomy, McGill University, Montr\'{e}al QC H3A-2T8.}

\author{L.-G.~Pang}
\affiliation{Department of Physics, University of California, Berkeley CA 94720.}
\affiliation{Nuclear Science Division, Lawrence Berkeley National Laboratory, Berkeley CA 94720.}

\author{J.~H.~Putschke}
\affiliation{Department of Physics and Astronomy, Wayne State University, Detroit MI 48201.}

\author{G.~Roland}
\affiliation{Department of Physics, MIT, Cambridge MA 02139.}

\author{S.~Rose}
\affiliation{Department of Physics and Astronomy, Texas A\&M University, College Station TX 77843.}
\affiliation{Cyclotron Institute, Texas A\&M University, College Station TX 77843.}

\author{B.~Schenke}
\affiliation{Department of Physics, Brookhaven National Laboratory, Upton NY 11973.}

\author{L.~Schwiebert}
\affiliation{Department of Computer Science, Wayne State University, Detroit MI 48202.}

\author{C.~Shen}
\affiliation{Department of Physics, Brookhaven National Laboratory, Upton NY 11973.}

\author{C.~Sirimanna}
\affiliation{Department of Physics and Astronomy, Wayne State University, Detroit MI 48201.}

\author{R.~A.~Soltz}
\affiliation{Lawrence Livermore National Laboratory, Livermore, CA 94550.}
\affiliation{Department of Physics and Astronomy, Wayne State University, Detroit MI 48201.}

\author{D.~Velicanu}
\affiliation{Department of Physics, MIT, Cambridge MA 02139.}

\author{G.~Vujanovic}
\affiliation{Department of Physics, The Ohio State University, Columbus OH 43210.}

\author{X.-N.~Wang}
\affiliation{Department of Physics, University of California, Berkeley CA 94720.}
\affiliation{Nuclear Science Division, Lawrence Berkeley National Laboratory, Berkeley CA 94720.}
\affiliation{Key Laboratory of Quark and Lepton Physics (MOE) and Institute of Particle Physics, Central China Normal University, Wuhan 430079, China.}

\author{R.~L.~Wolpert}
\affiliation{Department of Statistics, Duke University, Durham, NC 27708.}

\collaboration{The JETSCAPE Collaboration}

\date{\today}


\begin{abstract}

The modification of hard jets in an extended static medium held at a fixed temperature is studied using three different Monte-Carlo event generators (LBT, MATTER, MARTINI). Each event generator contains a different set of assumptions regarding the energy and virtuality of the partons within a jet versus the energy scale of the medium, and hence, applies to a different epoch in the space-time history of the jet evolution. For the first time, modeling is developed where a jet may sequentially transition from one generator to the next, on a parton-by-parton level, providing a detailed simulation of the space-time evolution of medium modified jets over a much broader dynamic range than has been attempted previously in a single calculation. Comparisons are carried out for different observables sensitive to jet quenching, including the parton fragmentation function and the azimuthal distribution of jet energy around the jet axis. The effect of varying the boundary between different generators is studied and a theoretically motivated criterion for the location of this boundary is proposed. The importance of such an approach with coupled generators to the modeling of jet quenching is discussed.

\end{abstract}

\maketitle


\section{Introduction}
\label{sec:introduction}

The extensive data sets and highly precise measurements that are now available for heavy-ion collisions at RHIC and the LHC, combined with theoretical developments, enable  an increasingly quantitative understanding of the properties of the Quark-Gluon Plasma (QGP)~\cite{Heinz:2015tua,Akiba:2015jwa}. The bulk behavior (related to low transverse momentum particles) of the plasma is described with  increasing precision using relativistic viscous fluid dynamics \cite{Romatschke:2007mq,Song:2010mg,Gale:2012rq,Heinz:2013th}; however, a detailed quantitative picture of the microscopic dynamics that is the cause of the evidently hydrodynamic behavior remains elusive. 

Jet modification \cite{Gyulassy:1993hr,Wang:1991xy,Baier:1998yf,Baier:1996sk,Baier:1996kr,Zakharov:1997uu,Zakharov:1996fv,Gyulassy:2003mc,Kovner:2003zj,CasalderreySolana:2007pr,Majumder:2010qh}, and its associated transport coefficients \cite{Baier:2002tc,Majumder:2008zg,Burke:2013yra} provide multi-scale hard probes of the QGP. Jets start with an off-shellness or virtuality $Q^2$ that is typically of the order of the hard scale, orders of magnitude higher than any scale in the medium. As the partons in a jet split and radiate more partons, they lose virtuality faster than they lose energy, and may spend a portion of their path in the medium at a scale comparable to the medium scale $Q^2 \sim \sqrt{\hat{q} E}$. Here, $E$ is the energy of the parton interacting with the medium, and $\hat{q}$ is the quark/gluon transport coefficient that denotes the broadening of the transverse (to the direction of the parton) momentum distribution, per unit length:
\begin{eqnarray}
\hat{q} = \frac{\langle p_\perp^2 \rangle_L}{L}.
\end{eqnarray}
In the equation above, $\langle \cdots \rangle_L$ denotes event average over a length $L$.
At even lower (or near thermal) momentum scales the jets are expected to be strongly coupled with the medium. 

The interaction of the jet with the medium and its splitting process differ at these various scales. To date, no calculation or simulation of jet modification in medium has accounted in a unified way for these different eras; all previous efforts have used a single formalism and applied it to the entire space-time history of the jet \cite{Burke:2013yra}. In this work, we provide the first treatment that accounts for different and complementary theoretical approaches, applied in succession to the space-time history of the jet in the medium. 

The energy of a reconstructed jet can be theoretically traced back to a single hard parton that undergoes subsequent splits and turns into a parton shower. The energy and virtuality of each parton determines its interaction with the medium. Hence, accurate event generators should allow for each parton to be ascribed the interaction formalism valid for that particular energy and virtuality at any given time. For instance, in the domain of high energy and high virtuality $Q^2 \gg \sqrt{ \hat{q} E}$, one expects to apply the medium modified Dokshitzer-Gribov-Lipatov-Altarelli-Parisi (DGLAP) evolution~\cite{Majumder:2009zu,Wang:2009qb} based on the higher-twist formalism~\cite{Wang:2001ifa,Majumder:2009ge}. As the virtuality approaches the medium induced scale, one expects to simulate a rate equation~\cite{Jeon:2003gi,Turbide:2005fk,Qin:2009bk} based on the 
BDMPS/AMY~\cite{Baier:1996sk,Baier:1996kr,Arnold:2001ms,Arnold:2002ja} formalisms or a derivative of the higher twist formalism~\cite{Wang:2013cia,Cao:2016gvr,Cao:2017hhk,Chen:2017zte}.

To carry out such a program requires detailed analysis of the effect of merging formalisms at the parton level. A full simulation of jet modification in realistic heavy-ion collisions requires several different components: an initial state calculation that samples hard partons from the parton distribution functions of the incoming nuclei and generates hard (high $Q^2$) collisions, leading to the formation of hard final state partons that interact with the medium, a viscous fluid dynamical simulation followed by hadronization of the soft sector, the modification of the jet as it propagates through this evolving medium, and hadronization of the jet partons. 

In this paper by the JETSCAPE Collaboration~\cite{JETSCAPE_website}, we perform the first such calculation of jet quenching, using coupled event generators that are applicable to different eras of the in-medium evolution of jet. We take a simplified approach in this study and consider the case of a single mono-energetic parton generated at the surface of a uniform, static medium of fixed length $L$ in the direction of propagation, with all chemical potentials set to zero and fixed temperature $T$. 
The medium is assumed to have an infinite transverse extent (a brick)~\cite{Armesto:2011ht}. This simplification of the problem has the advantage of eliminating possibly confounding effects from the dynamical evolution of the medium in a realistic heavy-ion collision, thereby focussing our attention on essential differences between the different energy loss formalisms applied to different jet evolution stages. There may also be effects that manifest themselves clearly in a static medium, e.g. the wake of a jet \cite{Neufeld:2009ep,Qin:2009uh}, but are blurred by a dynamical medium that undergoes phase transitions \cite{He:2015pra}. Comparison of simulations in the brick and in a dynamical medium will disentangle these different aspects of jet modification. 

In Ref.~\cite{Armesto:2011ht}, the TECHQM~\cite{techqm} and the JET collaboration~\cite{JET} studied four different energy loss formalisms within the brick setup. In that effort, calculations were carried out in the event-averaged formalism of energy loss of the leading parton, and all formalisms were treated separately. The goal of the current effort by the JETSCAPE Collaboration is a first comparison between different event generators for a full parton shower evolution within a brick, and to explore the effect of combining different generators at the partonic level. These results will serve as benchmarks for future event generators to be developed within the JETSCAPE framework. 

This paper is organized as follows: In Sec.~\ref{sec:methodology}, we provide a brief overview to the various simulation formalisms and describe how they are coupled together. In Sec.~\ref{sec:results}, we present the results of comparisons between the different approaches when used separately, along with those of the coupled approach. Concluding discussions and an outlook are provided in Sec.~\ref{sec:summary}.

\section{Methodology} 
\label{sec:methodology}

In this section, we first briefly summarize the three approaches for medium-modified parton showers to be employed here -- MATTER, MARTINI and LBT -- and then discuss how to combine the virtuality-ordered scheme for high virtuality partons and the time-ordered scheme for low virtuality partons into a unified theoretical approach. While MATTER represents the high virtuality scheme, MARTINI and LBT will be used for the low virtuality phase. 

It should be pointed out that the single gluon emission kernel, which is sampled by MARTINI and LBT, is quite different in terms of the physics assumptions used to derive it: MARTINI is based on the BDMPS/AMY formalism and LBT is based on the higher-twist formalism. Also, the LBT generator includes the dynamics of soft partons in the medium, an aspect missing in MARTINI (and MATTER) so far. Briefly stated, energy lost by hard partons in LBT reappears as an enhancement of the soft spectrum of partons, while energy lost by hard partons in MATTER and MARTINI is considered to be irrecoverably lost to the medium. As one would expect, the presence of a multitude of such soft particles leads to noticeable difference both in the soft spectrum and its angular distribution away from the jet axis.

\subsection{MATTER}
\label{subsec:MATTER}

The {\bf M}odular {\bf A}ll {\bf T}wist {\bf T}ransverse-scattering {\bf E}lastic-drag and {\bf R}adiation (MATTER) event generator simulates the splitting of high energy high virtuality jets, i.e. jets whose virtuality 
$Q^{2}  \gg \sqrt{\hat{q}E} $ where $E$ is the energy of the parton. At these high virtualities, the dominant mechanism of splitting is described by a medium-modified virtuality-ordered shower~\cite{Majumder:2009ge,Majumder:2009zu,Wang:2001ifa,Majumder:2011uk}. The underlying physical picture at this stage is that scattering in the medium produces a small variation in the vacuum shower process. The setup of the formalism ensures that the number of splittings dominates over the number of scatterings. 

A virtuality-ordered shower is initiated by a single hard parton produced at a point $r$, whose forward light-cone 
momentum $p^{+} = (p^0 + \hat{n}\cdot \vec{p} )/\sqrt{2}$ has been specified ($\hat{n} = \vec{p}/| \vec{p} |$ represents the direction of the jet). One then samples a Sudakov form factor to determine its virtuality $ t = Q^{2}$~\cite{Majumder:2013re,Majumder:2014gda}, 
\begin{eqnarray}
\Delta(t,t_{0}) &=& \exp \left[- \int\limits_{t_{0}}^{t}  \frac{dQ^{2}}{Q^{2}} \frac{\alpha_{s} (Q^{2})}{2\pi} \int\limits_{t_{0}/t}^{(1- t_{0}/t)}  dz P(z) \right. \\
& \times &
\left. \left\{ 1 +  \int\limits_{0}^{\zeta^{+}_{\mathrm{MAX}}}  d\zeta^{+} \frac{\hat{q} ( r + \zeta)  }{Q^{2} z(1-z)}  \Phi ( Q^{2}, p^{+}, \zeta^{+}  ) \right\} \right]. \nonumber
\end{eqnarray}
In the equation above, $\Phi$ represents a sum over phase factors that depend on $\zeta^+$, $p^+$, and $Q$. The transport coefficient $\hat{q}$ is evaluated at the 
location of scattering $\vec{r} + \hat{n} \zeta^{+}$. The function $P(z)$ is the 
vacuum splitting function. The maximum length sampled $\zeta^+_{\mathrm{MAX}}$ corresponds to $1.3 \tau_f^+$, where $\tau_f^+$ is the mean light-cone 
formation time $\tau^+_f = 2 p^+/Q^2$~\cite{Kordell:2017hmi}. 

Once $Q^2$ is determined, $z$ can be determined by sampling the splitting function $P(z)$. The transverse momentum of the produced pair 
(transverse to $\hat{n}$) is fixed once the shower is determined, by inspecting the difference in invariant mass between the parent and siblings of a
given split. To this one may add the transverse momentum generated by the propagation through the medium. This process is continued until the $Q^2$ reaches a predetermined value of $Q_0^2$. The parton splitting process stops at this point. The final partons at this stage may then be passed to a hadronization routine, or to another formalism. 

\subsection{MARTINI}
\label{subsec:MARTINI}

MARTINI ({\bf M}odular {\bf A}lgorithm for {\bf R}elativistic {\bf T}reatment of Heavy {\bf I}o{\bf N} {\bf I}nteractions) is a Monte Carlo event generator for simulating jets in heavy ion collisions~\cite{Schenke:2009gb}. The nucleon-nucleon collisions and the vacuum shower are generated by \textsc{Pythia} 8. 
Then MARTINI deals with the parton evolution in a QGP medium according to the AMY formalism for the radiative energy loss rates~\cite{Arnold:2002ja,Arnold:2002zm} combined with collisional processes~\cite{Schenke:2009ik}.
Those energy loss rates depend on the thermal background, whose information is provided by hydrodynamic calculations such as MUSIC~\cite{Schenke:2010nt}.
Finally, hadronization of the evolved partons can be performed by \textsc{Pythia} based on the Lund string model~\cite{Andersson:1983ia}. In this work, the initial parton shower is generated by MATTER, a static brick is adopted to study medium modification and all final spectra are analyzed at the parton level.

In MARTINI, the time evolution of the jet momentum distribution is governed by a set of coupled rate equations, which take the following forms:
\begin{align} 
\label{eq:rateMAR}
	\frac{dP_q(p)}{dt} &= \int\limits_k P_q(p+k)\frac{d\Gamma^{q}_{qg}(p+k,k)}{dkdt}
	- P_q(p)\frac{d\Gamma^{q}_{qg}(p,k)}{dkdt}\nonumber\\
	& +2P_g(p+k)\frac{d\Gamma^g_{q\bar{q}}(p+k,k)}{dkdt},\nonumber\\
	\frac{dP_g(p)}{dt} &= \int\limits_k P_q(p+k)\frac{d\Gamma^{q}_{qg}(p+k,p)}{dkdt}\\
	&+P_g(p+k)\frac{d\Gamma^{g}_{gg}(p+k,p)}{dkdt}\nonumber\\
	&-P_g(p)\left(\frac{d\Gamma^g_{q\bar{q}}(p,k)}{dkdt} + \frac{d\Gamma^g_{gg}(p,k)}{dkdt}\theta(2k-p)\right).\nonumber
\end{align}
$d\Gamma^a_{bc}(p,k)/dkdt$ is the transition rate for a process where a parton $a$ of energy $p$ emits a parton $c$ of energy $k$ and becomes a parton $b$.
The factor of 2 in front of $d\Gamma^g_{q\bar{q}}$ takes into account the fact that $q$ and $\bar{q}$ are distinguishable.
For the $g \rightarrow gg$ process, the $\theta$ function is there to avoid double counting of final states. Here $P_q(p)$ and
$P_g(p)$ are the energy distribution of quarks and gluons, respectively. The integration range with $k<0$ indicates energy gain from the thermal medium; the range with $k>p$ for $q \rightarrow qg$ process represents annihilation against anti-quark of energy $k-p$ from the medium.  

The AMY formalism describes energy loss of hard jets in heavy ion collisions as parton bremsstrahlung in the evolving QGP medium. 
The effective kinetic theory described in~\cite{Arnold:2002zm} assumes that quarks and gluons in the medium are well defined (hard) quasi-particles and have typical momentum of the order of temperature $T$ and thermal mass of order $gT$.
Under this assumption, the radiation rate can be calculated by means of integral equations~\cite{Arnold:2002ja}.

In the current version of MARTINI, the radiative energy loss mechanism is improved by implementing the effects of finite formation time and running coupling.
The formation time of the radiation process increases with $\sqrt{p}$ and a hard parton and an emitted parton are coherent within that time.
This interference effect suppresses the radiation rate at early times after the original radiation.

For the renormalization scale of running coupling constant $\alpha_s(\mu)$, we use the root mean square of the momentum transfer $\sqrt{\langle p^2_\perp \rangle}$ between the two particles, parameterized as
\begin{equation}
	\sqrt{\langle p^2_\perp \rangle} = (\hat{q}p)^{1/4},
\end{equation}
where $\hat{q}$ is the averaged momentum transfer squared per scattering and $p$ the energy of the mother parton \cite{Young:2012dv}.

\subsection{LBT}
\label{subsec:LBT}

A {\bf L}inear {\bf B}oltzmann {\bf T}ransport (LBT) model has been devised to describe the in-medium parton showers at low virtuality scale \cite{Li:2010ts,Wang:2013cia,He:2015pra,Cao:2016gvr,Cao:2017hhk,Chen:2017zte}. In the absence of a mean field, the evolution of the phase space distribution of a hard parton ``1" with $p_1^\mu = (E_1, \vec{p}_1)$ can be described using the Boltzmann equation
\begin{equation}
  \label{eq:boltzmann1}
  p_1\cdot\partial f_1(x_1,p_1)=E_1 (\mathcal{C}_\mathrm{el}+\mathcal{C}_\mathrm{inel}),
\end{equation}
in which $\mathcal{C}_\mathrm{el}$ and $\mathcal{C}_\mathrm{inel}$ are collision integrals for elastic and inelastic scatterings.

For elastic scattering, the collision term $\mathcal{C}_\mathrm{el}$ is evaluated with the leading-order matrix elements for all possible ``$12\rightarrow34$" processes between the given jet parton ``1" and a thermal parton ``2" present in the medium background. To regulate the collinear  ($u,t\rightarrow 0$) divergence of the matrix element, $S_2(s,t,u)=\theta(s\ge2\mu_\mathrm{D}^2)\theta(-s+\mu_\mathrm{D}^2\le t\le -\mu_\mathrm{D}^2)$ is imposed in which $\mu_\mathrm{D}^2=g^2T^2(N_c+N_f/2)/3$ is the Debye screening mass. The elastic scattering rate of parton ``1" can be evaluated as follows:
\begin{align}
 \label{eq:rate2}
 \Gamma_\mathrm{el}&=\sum_{2,3,4}\frac{\gamma_2}{2E_1}\int \frac{d^3 p_2}{(2\pi)^3 2E_2}\int\frac{d^3 p_3}{(2\pi)^3 2E_3}\int\frac{d^3 p_4}{(2\pi)^3 2E_4}\nonumber\\
&\times f_2(\vec{p}_2)\left[1\pm f_3(\vec{p}_3) \right]\left[1\pm f_4(\vec{p}_4)\right] S_2(s,t,u)\nonumber\\
&\times (2\pi)^4\delta^{(4)}(p_1+p_2-p_3-p_4)|\mathcal{M_\mathrm{12\rightarrow34}}|^2,
\end{align}
in which $\gamma_2$ represents the spin-color degeneracy of parton ``2". And therefore, the probability of elastic scattering of parton ``1" in each time step $\Delta t$ is $P_\mathrm{el}=\Gamma_\mathrm{el}\Delta t$. 

For inelastic scattering, or medium-induced gluon radiation, the average number of emitted gluons from a hard parton in each time step $\Delta t$ is evaluated as \cite{Cao:2013ita,Cao:2015hia,Cao:2016gvr}
\begin{equation}
 \label{eq:gluonnumber}
 \langle N_g \rangle(E,T,t,\Delta t) = \Delta t \int dxdk_\perp^2 \frac{dN_g}{dx dk_\perp^2 dt},
\end{equation}
in which the differential spectrum of radiated gluon is taken from the higher-twist energy loss formalism \cite{Guo:2000nz,Majumder:2009ge,Zhang:2003wk}:
\begin{eqnarray}
\label{eq:gluondistribution}
\frac{dN_g}{dx dk_\perp^2 dt}=\frac{2\alpha_s C_A \hat{q} P(x)k_\perp^4}{\pi \left({k_\perp^2+x^2 m^2}\right)^4} \, {\sin}^2\left(\frac{t-t_i}{2\tau_f}\right),
\end{eqnarray}
where $x$ and $k_\perp$ are the fractional energy and transverse momentum of the emitted gluon with respect to its parent parton, $\alpha_s$ is the strong coupling constant, $C_A=N_c$ is the gluon color factor, $P(x)$ is the splitting function, and $\hat{q}$ is the transport coefficient due to elastic scattering and can be obtained by evaluating Eq.~(\ref{eq:rate2}) weighted by the transverse momentum broadening of parton ``1". The mass dependence of gluon emission from heavy quark is included in Eq.~(\ref{eq:gluondistribution}). In addition, $t_i$ denotes an ``initial time" or the production time of the parent parton from which the gluon is emitted, and $\tau_f={2Ex(1-x)}/{(k_\perp^2+x^2m^2)}$ is the formation time of the radiated gluon. To avoid possible divergence as $x\rightarrow 0$, a lower cut-off $x_\mathrm{min}=\mu_D/E$ is implemented for the energy of emitted gluon. Multiple gluon radiation is allowed in each time step. Different emitted gluons are assumed independent with each other and therefore their number $n$ obeys a Poisson distribution with the mean as $\langle N_g \rangle$:  
\begin{eqnarray}
\label{eq:possion}
P(n)=\frac{\langle N_g\rangle^n}{n!}e^{-\langle N_g\rangle^n}.
\end{eqnarray}
Thus, the probability for the total inelastic scattering process is $P_\mathrm{inel}=1-e^{-\langle N_g \rangle}$. Note that for the $g\rightarrow gg$ process, $\langle N_g\rangle/2$ is taken as the mean instead to avoid double counting.

To combine the above elastic and inelastic processes, the total scattering probability is divided into two regions: pure elastic scattering with probability $P_\mathrm{el}(1-P_\mathrm{inel})$ and inelastic scattering with probability $P_\mathrm{inel}$. Thus the total scattering probability is $P_\mathrm{tot}=P_\mathrm{el}+P_\mathrm{inel}-P_\mathrm{el}\cdot P_\mathrm{inel}$. Based on these probabilities, the Monte Carlo method is applied to determine whether a given jet parton is scattered inside the thermal medium and whether the scattering is pure elastic or inelastic. With a selected scattering channel, the energies and momenta of the outgoing partons are then sampled based on the corresponding differential spectra given by Eq.~(\ref{eq:rate2}) and (\ref{eq:gluondistribution}). The only parameter in this LBT model is the strong coupling constant $\alpha_s$ that quantifies the jet-medium interaction, which is determined by comparing model calculation to experimental data of single heavy and light flavor hadron production, single inclusive jet production and $\gamma$-jet production in heavy-ion collisions \cite{Cao:2017hhk,Cao:2016gvr,Li:2010ts,Wang:2013cia,Chen:2017zte}.

\subsection{Combining MATTER and MARTINI/LBT}
\label{subsec:combine}

To establish a unified framework for parton showers, we apply MATTER to partons with large virtuality and MARTINI or LBT to partons with low virtuality. MARTINI and LBT, based on AMY and higher-twist energy loss formalisms respectively, will be compared to each other and shown to be consistent within the kinematic region we investigate. To begin with, each hard parton directly produced in hard scatterings is placed in MATTER, in which its virtuality-ordered splitting process is simulated as described in Sec.~\ref{subsec:MATTER}. In each splitting, the virtualities of the daughter partons are much smaller than that of the parent. If the virtuality of a given parton in the shower drops below a certain scale $Q_0$, it is passed to MARTINI/LBT for the subsequent time-ordered evolution inside the medium. 

One crucial quantity in this unified framework is the separation scale $Q_0$ between MATTER and MARTINI/LBT evolutions. Two different schemes, fixed $Q_0$ and dynamical $Q_0$, are applied and compared in this study. For the former, fixed values of $Q_0$ (1, 2 or 3 GeV) are used as the scale below which both vacuum and medium-modified showers in MATTER cease. For the latter, we define $Q_0^2=\hat{q}\tau_f$ for each parton. This quantifies the average virtuality gain of each parton from its scattering with the medium and serves as a reasonable separation scale between the virtuality-loss splitting process in MATTER and the near-constant-virtuality scattering process in MARTINI/LBT. With $\tau_f=2E/Q_0^2$, we have
\begin{eqnarray}
\label{eq:Q02}
Q_0^2=\sqrt{2E\hat{q}},
\end{eqnarray}
in which the quark/gluon transport coefficient is obtained from a finite temperature field theory calculation as \cite{Gyulassy:1993hr,CaronHuot:2010bp}, 
\begin{eqnarray}
\label{eq:qhat}
\hat{q}=C_R\alpha_s\mu_D^2 T \log\left(\frac{6ET}{\mu_D^2}\right).
\end{eqnarray}
For small $E$, $Q_0=1$~GeV is used if the estimate above yields a $Q_0$ smaller than 1~GeV.
Thus $Q_0$ is dynamically determined based on the energy of each parton and the local temperature of its surrounding medium. 

Note that for this dynamical scenario, Eqs.~(\ref{eq:Q02},\ref{eq:qhat}) are only well defined in a thermal medium, and thus fixed $Q_0=1$~GeV is used when partons travel outside the color deconfined nuclear matter (or in vacuum). The value $Q_0=1$~GeV is from the standard practice to regard 1~GeV to be the lowest scale where pQCD is expected to be valid. Partons with a virtuality below this scale are assumed to be strongly coupled with the medium, or undergo hadronization if outside the QGP. Since neither of these effects are incorporated in this first attempt at a multi-stage event generator, the $Q_0$ is held at a minimum of 1~GeV until exit from the brick.

In this work, we start with a single quark and let it evolve either in vacuum or through a static brick with fixed temperature 250~MeV. With a virtuality greater than $Q_0$, each parton in the shower evolves in MATTER; but after hitting $Q_0$, it starts evolving in either MARTINI or LBT. We will investigate how such combined theoretical approach affects the energy distribution and jet broadening of the finally produced partons, compared to traditional methods in the literature where a single energy loss approach is applied. The initial energy of the single quark and the length of the brick medium will be varied so that one may study in which region the virtuality ordered splitting in the parton shower is more prevalent and in which region the time ordered splitting dominates. Note that it is possible that a highly virtual parton still has not hit $Q_0$ after it traverses the entire brick medium. If so, the vacuum shower is attached in MATTER after its in-medium evolution until each daughter parton reaches $Q_0$; and MARTINI/LBT does not have any effect on the shower in such case. Therefore, all plots describe the distribution of partons at the exit from the brick or the moment hitting $Q_0$, whichever comes later. Throughout our calculation, a fixed value of $\alpha_s=0.3$ is used in the MARTINI/LBT portion and 
a running $\alpha_s$ used in the higher virtuality MATTER portion.

\begin{figure}[tb]
  \epsfig{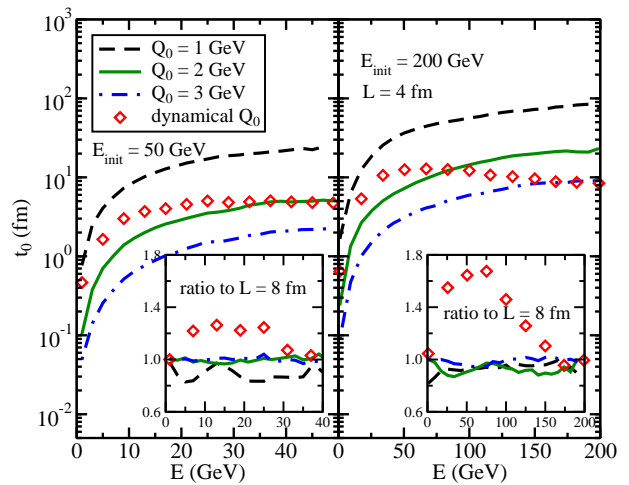}
  \caption{(Color online) Separation time $t_0$ between MATTER and MARTINI/LBT as functions of parton energy, for different initial parton energy $E_\mathrm{init}$, brick size $L$ and separation virtuality scale $Q_0$. The main figures show results in a brick with length of 4~fm, and the inset subfigures show the ratio between 4~fm and 8~fm. 
}
 \label{fig:plot-t0vsE}
\end{figure}

Before discussing the energy distribution and jet broadening of the final partons, we first investigate the time $t_0$ at which a parton in the shower hits $Q_0$ as a function of its energy in Fig.~\ref{fig:plot-t0vsE}. This $t_0$ is obtained by summing the formation times of all previous splittings in MATTER before the produced parton hit $Q_0$ as discussed in Sec.~\ref{subsec:MATTER}. In Fig.~\ref{fig:plot-t0vsE}, the left panel corresponds to 50~GeV energy for the initial quark, and the right panel corresponds to 200~GeV. In each panel, 3 different fixed $Q_0$ and the dynamical $Q_0$ scenarios are compared. The two main figures show results in a brick with length of 4~fm, and the two subfigures inside show the ratio between 4~fm and 8~fm. 

From these figures, one may observe the time it takes to evolve an energetic parton down to $Q_0$ can be long (compared to the formation time of the QGP $\tau_0\sim0.6$~fm in realistic heavy-ion collisions). The switching time $t_0$ increases if the separation scale $Q_0$ decreases or the initial energy $E_\mathrm{init}$ (i.e. the possible maximum virtuality) of the parton increases. For $E_\mathrm{init}=50$~GeV, $t_0$ for the dynamical $Q_0$ is consistent with that for fixed $Q_0=2$~GeV at the high energy end of the final parton spectrum. This is a natural result from Eq.~(\ref{eq:Q02}). And if $E_\mathrm{init}$ is increased to 200~GeV, $t_0$ for the dynamical $Q_0$ is then consistent with that for fixed $Q_0=3$~GeV at the high energy end. On the other hand, for final partons with lower energies, $t_0$ for the dynamical $Q_0$ approaches that for fixed $Q_0=1$~GeV since at low $E$ and after partons travel outside the brick (back into vacuum) $Q_0$ is set as 1~GeV. 

Changing the size of the brick from $L=4$ to 8~fm also affects the value of $t_0$. For the scenarios of fixed $Q_0$, extending the length of the brick increases scattering of partons inside the medium. This is a virtuality gain process and thus may delay the time $t_0$ for each parton to hit $Q_0$. For $E_\mathrm{init}=50$~GeV, this only affects the $Q_0=1$~GeV scenario, since for $Q_0=2$ and 3~GeV, most partons hit $Q_0$ before 4~fm and thus adding another 4~fm of length/time has slight effect. On the contrary, for $E_\mathrm{init}=200$~GeV, extending $L$ from 4 to 8~fm clearly increases $t_0$ for both fixed $Q_0=1$ and 2~GeV. The opposite effect of varying the brick size is observed for the scenario of dynamical $Q_0$. Unlike the fixed $Q_0$ scenario where the same $Q_0$ is adopted for both in-medium and vacuum parton shower, the dynamical scenario uses $Q_0$ from Eq.~(\ref{eq:Q02}) (usually larger than 1~GeV) inside the brick but 1~GeV outside. And therefore, extending $L$ from 4 to 8~fm increases the range where larger $Q_0$ is applied and thus shortens $t_0$. These discussions on the separation time $t_0$ between MATTER and MARTINI/LBT evolutions will be helpful in understanding the final parton spectra within our unified theoretical approach as will be shown in the next section.

\section{Energy distribution and jet broadening in MATTER+MARTINI/LBT}
\label{sec:results}

In this section, we will present the energy distribution and jet angular distribution of the parton shower and compare our unified approach to traditional approaches in which a single energy loss scheme is applied. In the simplified scheme of the brick, these two distributions form the underlying basis of almost all jet observables. In the limit of near onshell particles, only three components of the four-momentum are relevant, which we have chosen as the energy $E_{k}$, and the components transverse to the jet axis, 
$k_\perp$ [$k = \left( E_{k}^{2} - Q_{0}^{2} - k_{\perp}^{2}  \right)$, where $Q_0 =1$~GeV].

\begin{figure}[tb]
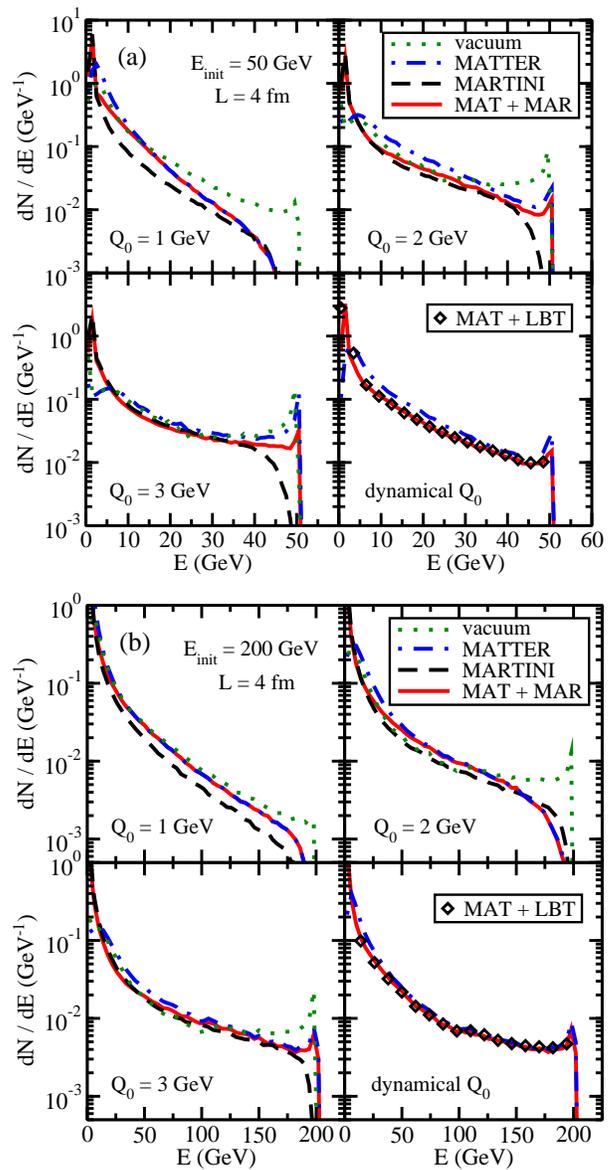

 \subfigure{\label{fig:plot-dNdE-E050L04}
  \epsfig{file=plot-dNdE-E050L04.eps, width=0.44\textwidth, clip=}}
 \subfigure{\label{fig:plot-dNdE-E200L04}
  \epsfig{file=plot-dNdE-E200L04.eps, width=0.44\textwidth, clip=}}
  \caption{(Color online) Energy distribution of final partons with a $L=4$~fm brick: (a) for $E_\mathrm{init}$=50 GeV and (b) for 200~GeV for the initial quark.}
  \label{fig:plot-dNdE-L04}
\end{figure}

\begin{figure}[tb]
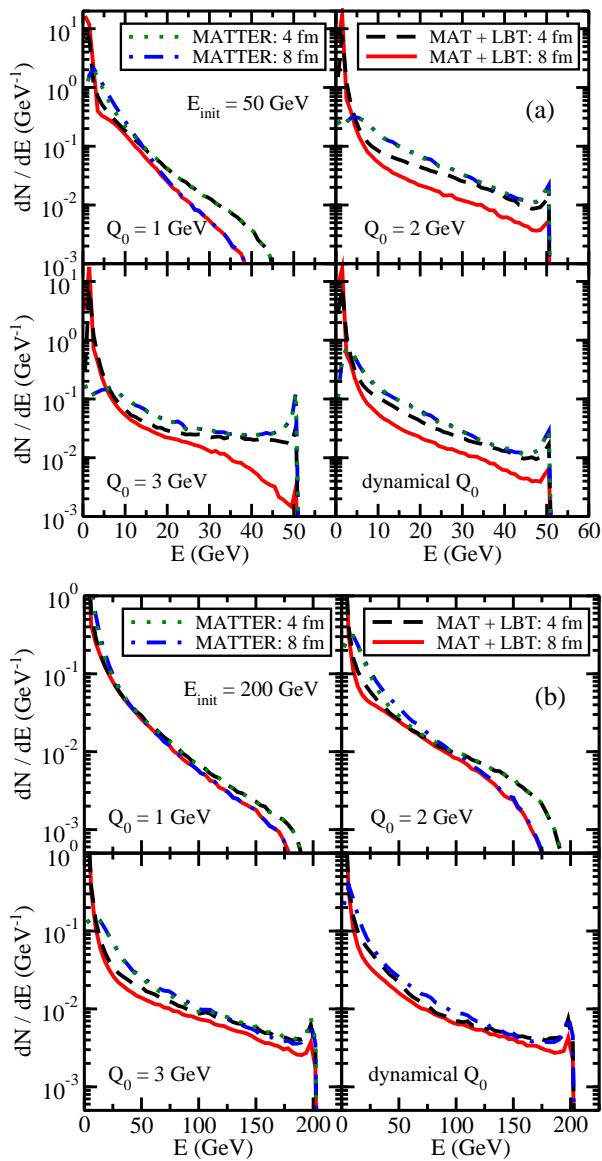

 \subfigure{\label{fig:plot-dNdE-E050LDep}
  \epsfig{file=plot-dNdE-E050LDep.eps, width=0.44\textwidth, clip=}}
 \subfigure{\label{fig:plot-dNdE-E200LDep}
  \epsfig{file=plot-dNdE-E200LDep.eps, width=0.44\textwidth, clip=}}
  \caption{(Color online) The length dependence of the energy spectra: (a) for $E_\mathrm{init}$=50 GeV and (b) for 200~GeV for the initial quark.}
  \label{fig:plot-dNdE-LDep}
\end{figure}

In Fig.~\ref{fig:plot-dNdE-L04}, we show the energy distribution of final partons with the brick size $L=4$~fm. The parton energy of 50~GeV is used for the initial quark in Fig.~\ref{fig:plot-dNdE-E050L04} and 200 GeV for Fig.~\ref{fig:plot-dNdE-E200L04}. In each figure, the four panels correspond to different choices of the separation scale $Q_0$: three fixed cases and one dynamical case. For the three panels with fixed $Q_0$, we compare results between vacuum shower, medium modified shower through MATTER alone, MARTINI alone and MATTER+MARTINI. For the panel with dynamical $Q_0$, we compare results between medium modified shower through MATTER alone, MATTER+MARTINI and MATTER+LBT. Note that for the vacuum shower or the medium modified shower in MATTER alone, we let all partons evolve down to the given $Q_0$ in the virtuality ordered scheme. For the pure MARTINI evolution, we let each parton produced by the vacuum shower evolve through MARTINI for the entire 4 fm as the traditional implementation in the literature. MATTER+MARTINI/LBT corresponds to our unified approach as discussed in Sec.~\ref{subsec:combine} in which each parton evolves though MATTER till $t_0$ (i.e., with virtuality larger than $Q_0$) after which it evolves through MARTINI or LBT until 4~fm.

From Fig.~\ref{fig:plot-dNdE-E050L04}, we observe the energy spectra of final partons from MARTINI and LBT are consistent with each other, within our setup, except for the very low energy region, since in this calculation MARTINI is set up to regard emitted partons with an energy less than 2 GeV as a part of the medium. The similarity between MARTINI and LBT may be due to the fact that the results shown in Fig.~\ref{fig:plot-dNdE-L04} are consequence of multiple emissions. Even though the single gluon emission spectra differ in MARTINI and LBT in detail, the multiple emission spectra obtained either by solving Eq.~(\ref{eq:rateMAR}) or via the Poisson ansatz Eq.~(\ref{eq:possion}) may wash away some of the differences.

For $E_\mathrm{init}=50$~GeV, the maximum scale of the medium $\sqrt{\hat{q}\tau_f}$ probed by the hard parton is around 2~GeV, and therefore pure MATTER evolution (compared to the vacuum shower) leads to a minimal suppression of the spectra if one sets $Q_0\gtrsim2$~GeV. Within our unified approach of MATTER+MARTINI, we see that the effect of just applying MARTINI for the entire length of the medium on a vacuum shower leads to a considerably larger suppression of the spectrum of final state partons, specifically for high energy partons, compared to the suppression of first allowing the partons to split using MATTER and then with MARTINI once the virtuality reaches $Q_0$.
The cause of this, clearly illustrated in Fig.~\ref{fig:plot-t0vsE},  is that it takes a longer time for more energetic partons to evolve down to $Q_0$ and therefore leaves shorter time for them to evolve inside MARTINI. This is more apparent when $Q_0$ is smaller. In the end, results for the dynamical $Q_0$ scenario is close to the fixed $Q_0=2$~GeV scenario which can be naturally understood from Eq.~(\ref{eq:Q02}).

If the energy of the initial quark is increased by a factor of 4, one may observe in Fig.~\ref{fig:plot-dNdE-E200L04} that the amount of suppression obtained by pure MATTER evolution (vs. vacuum) is considerable up to a higher $Q_0$ cut (around 3~GeV), since the scale of the medium probed by the jet increases with the jet energy. In addition, the additional amount of suppression obtained from the MARTINI portion in our MATTER+MARTINI approach is reduced compared to the $E_\mathrm{init}=50$~GeV scenario, since the switching time $t_0$ between MATTER and MARTINI is larger (see Fig.~\ref{fig:plot-t0vsE}). For $E_\mathrm{init}= 200$~GeV, results with dynamical $Q_0$ are closer in form to those with the fixed $Q_0=3$~GeV. 

In Fig.~\ref{fig:plot-dNdE-LDep} we investigate the path length dependence of the energy distribution of the final partons within our MATTER+LBT framework. In Fig.~\ref{fig:plot-dNdE-E050LDep}, the energy of the initial quark is set as 50~GeV. We observe for the case of pure MATTER evolution with $Q_0=1$~GeV, the energy loss of the incoming parton is greater, i.e., the spectra is more suppressed, when the medium size $L$ is extended from 4 to 8~fm. However, for $Q_0\gtrsim2$~GeV, there is no apparent difference between $L=4$ and 8~fm since most partons evolve down to the virtuality of $Q_0$ before 4~fm (see Fig.~\ref{fig:plot-t0vsE}) and these partons are propagated without effect through the remaining  4~fm of medium. On the other hand, extending $L$ from 4 to 8 fm leaves longer time for parton evolution through LBT. In other words, the LBT (and MARTINI) evolution has a much larger effect on the suppression if the path length inside the medium is longer, as is readily observed for $Q_0\gtrsim2$~GeV. For $Q_0=1$~GeV, the separation point between the pure MATTER curve and the MATTER+LBT curve also shifts to the right (from around 10 to around 20 GeV) when $L$ is extended from 4 to 8~fm, indicating a wider range of partons affected by the LBT evolution when $L$ is larger.

A similar investigation of the length dependence is presented in Fig.~\ref{fig:plot-dNdE-E200LDep} where the energy of the initial quark is increased to 200~GeV. Compared to Fig.~\ref{fig:plot-dNdE-E050LDep}, the difference between $L=4$ and 8~fm for the pure MATTER scenario is not only obvious for $Q_0=1$~GeV, but also for $Q_0=2$~GeV now, because with $E_\mathrm{init}=200$~GeV, it takes much longer than 4~fm for MATTER to evolve partons down to 2~GeV (see Fig.~\ref{fig:plot-t0vsE}) and thus evolution between 4 and 8~fm becomes important. However, since now it takes longer for MATTER to evolve partons down to a given $Q_0$, shorter time is left for the subsequent LBT evolution and thus extending $L$ from 4 to 8~fm results in a lower effect of the LBT contribution to the suppression of the final parton spectra compared to the previous results for $E_\mathrm{init}=50$~GeV.

\begin{figure}[tb]
  \epsfig{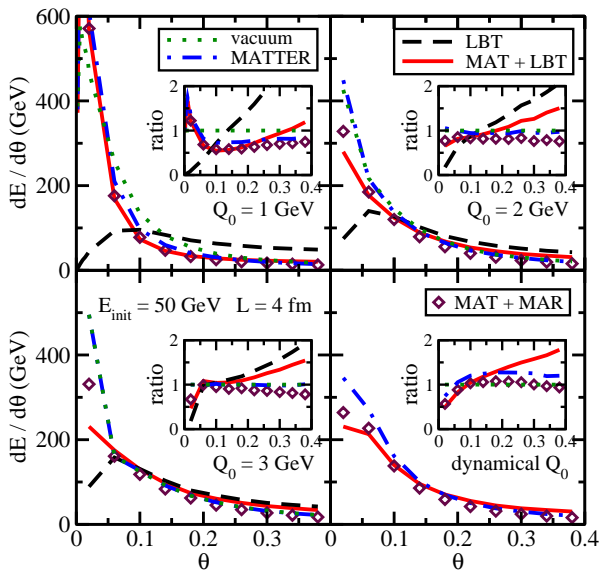}
  \caption{(Color online) Energy distribution with respect to the jet cone angle (for $E_\mathrm{init}=50$~GeV and $L=4$~fm).}
 \label{fig:plot-dEdth-E050L04}
\end{figure}

Finally, we also investigate the energy flow inside the jet within our unified approach of MATTER+MARTINI/LBT. In Fig.~\ref{fig:plot-dEdth-E050L04}, we present the energy distribution with respect to the jet cone angle, compared between vacuum shower, in-medium shower from pure MATTER, pure LBT and MATTER+LBT/MARTINI. The four panels are for the four different choices of $Q_0$, and in order to clearly display the effect of medium modification, we show the ratios between the medium modified spectra and their corresponding vacuum spectra in the four subfigures (For the dynamical scenario, $Q_0=1$~GeV is used for the vacuum baseline in calculating the ratio). 

In most cases, medium induced emission leads to a depletion of the energy at small angles and a milder enhancement at larger angles. 
However, there are several caveats in this effect. If the switching virtuality is set at $Q_{0} = 1$~GeV, or MATTER evolution is allowed to run down to the minimum possible value, then there is a narrowing of the cone (enhancement) at very small angles ($\theta < 0.05$). This is mostly caused by MATTER evolution.  
Both MARTINI and LBT lead to a suppression of the energy in the most collinear bins.

In MATTER, the narrowing is caused by the enhanced Sudakov factor, which leads to 
an enhanced probability of collinear emission. 
The broadening and enhancement at larger angles is mostly present in LBT. The particles shifted to larger angles tend to be low energy partons, which are explicitly excised from the MATTER and MARTINI showers. 
Also, LBT is more effective in shifting the energy distribution into larger angles than MATTER \cite{Majumder:2005sw}, since elastic scattering is also included in LBT. The contributions from LBT and MARTINI are also compared: MARTINI is consistent with LBT at small angle, or for high-energy partons, but they deviate at large angles since partons with very low energy are regarded as part of the background medium in the current MARTINI simulation. 

The results presented in Fig.~\ref{fig:plot-dEdth-E050L04} are somewhat surprising, specifically those in the top left panel, where $Q_0 = 1$~GeV. The multi-stage simulation involving MATTER and LBT qualitatively reproduces the feature seen in experimental data of an enhancement at small angles, a depletion at intermediate, and an enhancement at larger angles (see inset). 
This is a feature absent in any single generator applied to a static medium. It has been proposed that a back reaction from a dynamical medium, enhanced by radial flow is responsible for the enhancement at large angles~\cite{Casalderrey-Solana:2016jvj}. The simulations presented in this work, do not involve a dynamical medium, however, they do involve elastic scattering of soft partons off the medium. Due to the switch between MATTER and LBT at $Q_0 = 1$~GeV, hard partons at the lowest angles spend most of their lifetime in the MATTER phase and are not affected by the rescattering. Whereas, soft wide angle partons that reach a virtuality of $Q_0$ deep in the medium, are scattered out to larger angles, this leads to the depletion at intermediate angles and the enhancement at larger angles. As $Q_0$ is increased, even the hard partons reach the transition between MATTER and LBT some distance from exit and are affected by the re-scattering in LBT. As a result, the enhancement at small angles is lost.

\section{Summary}
\label{sec:summary}

In this work, different event generators for medium modified parton showers are coupled for the first time at the parton level. The virtuality-ordered event generator MATTER, based on the higher twist energy loss formalism, is adopted for shower partons at large virtuality, while two time-ordered transport models, MARTINI and LBT, are applied at low virtuality. MARTINI, based on the AMY energy loss formalism, and LBT, based on the higher twist energy loss formalism, are shown to be consistent with each other. Both fixed and dynamical separation scales between MATTER and MARTINI/LBT have been explored in this study. Varying the parton energy, medium length and virtuality separation scale $Q_0$, we studied the changing relative weights of the different energy loss schemes on the medium modifications experienced by the jet.  

It was shown that the time $t_0$ it takes to evolve each parton down to a given $Q_0$ can be large. The switching time $t_0$ grows with increasing initial and final parton energies and decreases with larger $Q_0$. Its dependence on the size of the thermal medium varies with the parton energy and $Q_0$. When combining these two schemes -- MATTER and MARTINI/LBT -- into a unified approach, we observe that MATTER plays a larger role in the evolution of high energy partons whereas MARTINI/LBT play a larger role for lower energy ones since the latter reach $Q_0$ earlier and thus are subject to MARTINI/LBT evolution for a larger fraction of time. A larger value of $Q_0$ typically suppresses the medium modification by MATTER and thus increases the relative contribution from MARTINI/LBT. The importance of the MARTINI/LBT contribution also increases with the path length in the medium, but this increase may be suppressed if the initial parton starts with a higher energy. Finally, it was shown that jet can be broadened more effectively, i.e. stronger energy flow towards larger jet cone angles, if elastic scattering is included in addition to parton splitting. And combining MATTER and LBT may provide the non-monotonic nuclear modification of the angular distribution of jet energy within a small jet cone -- a feature that is hard to obtain when a single energy loss mechanism is applied.

This work contributes a crucial step towards establishing a complete theoretical picture of parton shower evolution in relativistic heavy-ion collisions, applying within a single unified framework different energy loss formalisms based on complementary approximations, consistently to different kinematic regions during the parton evolution history. Our study will be extended in the near future in two directions. The framework will be coupled to a hydrodynamic background in order to study jet evolution in a realistically evolving dynamical medium. Secondly, additional theoretical schemes for parton evolution will be added to this framework, such as the AdS/CFT formalism for almost thermalized partons. All these insights will be incorporated in the upcoming JETSCAPE framework. 


\section*{Acknowledgments} 
This  work  was  supported  in  part  by  the National Science Foundation (NSF) within the framework of the JETSCAPE collaboration, under grant numbers ACI-1550172 (R.A.B. and D.V.), ACI-1550221(S.R. and R.J.F.), ACI-1550223 (D.B. and U.H.), ACI-1550225 (J.B. and J.C.), ACI-1550228 (L.-G.P. and X.-N.W.), and ACI-1550300 (S.C., A.K., D.P., A.M., C.Si. and R.A.S.). It was also supported in part by the NSF under grant numbers PHY-1207918 (M.K.), 1516590 (S.R. and R.J.F.), and by the US Department of Energy, Office of Science, Office of Nuclear Physics under grant numbers \rm{DE-AC02-05CH11231} (Y.H. and X.-N.W.), \rm{DE-AC52-07NA27344} (R.A.S.), \rm{DE-SC0012704} (B.S. and C.S.), \rm{DE-SC0013460} (S.C., A.K., A.M. and C.Si.), \rm{DE-SC0004286} (U.H.), and \rm{DE-FG02-05ER41367} (S.A.B.). The work was also supported in part by the National Science Foundation of China (NSFC) under grant number 11521064, Ministry of Science and Technology (MOST) of China under Projects number 2014CB845404 (Y.H., T.L. and X.-N.W), and in part by the Natural Sciences and Engineering Research Council of Canada (C.G., S.J. and C.P.) and the Fonds de recherche du Qu\'ebec - Nature et technologies (FRQ-NT) (G.V.).  Computations were made in part on the supercomputer \emph{Guillimin} from McGill University, managed by Calcul Qu\'ebec and Compute Canada. The operation of this supercomputer is funded by the Canada Foundation for Innovation (CFI), NanoQu\'ebec, R\'eseau de M\'edicine G\'en\'etique Appliqu\'ee~(RMGA) and FRQ-NT. C.G. gratefully acknowledges support from the Canada Council for the Arts through its Killam Research Fellowship program. C.S. gratefully acknowledges a Goldhaber Distinguished Fellowship from Brookhaven Science Associates.


\bibliography{SCrefs}

\end{document}